\documentclass[lnicst]{svmultln}
\usepackage{makeidx}  

\usepackage{caption}
\usepackage{graphicx}
\usepackage{url}
\usepackage[autostyle]{csquotes}
\usepackage{amssymb,amsmath}
\usepackage{epsfig}
\usepackage{epstopdf}
\usepackage{setspace}
\usepackage{amsthm}
\usepackage{enumerate}
\usepackage{tikz}
\usetikzlibrary{shapes,arrows}
\usepackage{algorithm}
\usepackage{algcompatible}
\usepackage{float}
\usepackage{algpseudocode}
\usepackage{pifont}
\newtheorem{assumption}{Assumption}

\begin{document}
	\mainmatter              
	\title{Nash Equilibrium Seeking with Non-doubly Stochastic Communication Weight Matrix}
	\titlerunning{Nash Equilibrium Seeking}  
	%
	\author{Farzad Salehisadaghiani \and Lacra Pavel\thanks{This work was supported by an NSERC Discovery Grant.}}
	\authorrunning{F. Salehisadaghiani and L. Pavel}   
	%
	%
	\institute{Department of Electrical and Computer Engineering, University of Toronto, 10 King's College Road, Toronto, ON, M5S 3G4, Canada
		\email{farzad.salehisadaghiani@mail.utoronto.ca}, \email{pavel@control.utoronto.ca}}
	
	\maketitle              
	
	\begin{abstract}
		A distributed Nash equilibrium seeking algorithm is presented for networked games. We assume an incomplete information available to each player about the other players' actions. The players communicate over a strongly connected digraph to send/receive the estimates of the other players' actions to/from the other local players according to a gossip communication protocol. Due to asymmetric information exchange between the players, a non-doubly (row) stochastic weight matrix is defined. We show that, due to the non-doubly stochastic property, the total average of all players' estimates is not preserved for the next
		iteration which results in having no exact convergence. We present an almost sure convergence proof of the algorithm to a Nash equilibrium of the game. Then, we extend the algorithm for graphical games in which all players' cost functions are only dependent on the local neighboring players over an interference digraph. We design an assumption on the communication digraph such that the players are able to update all the estimates of the players who interfere with their cost functions. It is shown that the communication digraph needs to be a superset of a transitive reduction of the interference digraph. Finally, we verify the efficacy of the algorithm via a simulation on a social media behavioral case.
	\end{abstract}
	\section{Introduction}
	The problem of finding a Nash equilibrium (NE) of a networked game has recently drawn many attentions. The players who participate in this game aim to minimize their own cost functions selfishly by making decision in response to other players' actions. Each player in the network has only access to local information of the neighbors. Due to the imperfect information available to players, they maintain an estimate of the other players' actions and communicate over a communication graph in order to exchange the estimates with local neighbors. Using this information, players update their actions and estimates.
	
	In many algorithms in the context of NE seeking problems, it is assumed that the communications between the players are symmetric in the sense that the players who are in communication can exchange their information altogether and update their estimates at the same time. This, in general, leads to a doubly stochastic communication weight matrix which preserves the global average of the estimates over time. However, there are many real-world applications in which symmetric communication is not of interest or may be an undesired feature in applications such as sensor network.\\
	\emph{Literature review.} Our work is related to the literature on Nash games and distributed NE seeking algorithms \cite{jayash8,yin2011nash,frihauf2012nash,jayash11,salehisadaghiani2016distributedifac}. A distributed algorithm is proposed in \cite{zhu2016distributed} to compute a generalized NE of the game for a complete communication graph. In \cite{Jayash}, an algorithm is provided to find an NE of \emph{aggregative games} for a partial communication graph but complete interference graph. This algorithm is extended by \cite{salehisadaghiani2016distributed} for a more general class of games in which the players' cost functions are not necessarily dependent on the aggregate of the players' actions. It is further generalized to the case with partial interference graph in \cite{ssalehisadaghiani2016distributed}.  For a \emph{two-network zero-sum game} \cite{gharesifard2013distributed} considers a distributed algorithm for NE seeking. To find distributed algorithms for games with local-agent utility functions, a methodology is presented in \cite{Marden2013} based on state-based \emph{potential games}.
	
	Gossip-based communication has been widely used in synchronous and asynchronous algorithms in consensus and distributed optimization problems \cite{nedic2011asynchronous,aysal2009broadcast,fagnani2008randomized}. In \cite{nedic2011asynchronous}, a gossip algorithm is designed for a distributed broadcast-based optimization problem. An almost-sure convergence is provided due to the non-doubly stochasticity of the communication matrix. In \cite{aysal2009broadcast}, a broadcast gossip algorithm is studied to compute the average of the initial measurements which is proved to converge almost surely to a consensus.\\
	\emph{Contributions.} We propose an asynchronous gossip-based algorithm to find an NE of a distributed game over a communication digraph. We assume that players send/receive information to/from their local out/in-neighbors over a strongly connected communication digraph. Players update their own actions and estimates based on the received information. We prove an almost sure convergence of the algorithm to the NE of the game. {\emph{Unlike in the undirected case \cite{salehisadaghiani2016distributed,ssalehisadaghiani2016distributed}, herein we cannot exploit the doubly stochastic property for the communication weight matrix due to asymmetric information exchange. Non-doubly stochastic property leads to have total average of the players' estimates not preserved over time. This was one of the critical steps in the
			convergence proof in \cite{salehisadaghiani2016distributed,ssalehisadaghiani2016distributed}.}}
	
	Moreover, we extend the algorithm for graphical games in which the players' cost functions may be interfered by any subset of players' (not necessarily all the players') actions. The locality of cost functions is specified by an interference digraph which marks the pair of players who interfere one with another. In order to have a convergent algorithm, we design an assumption on the communication digraph by which there exists a lower bound on the communication digraph which is a  transitive reduction of the interference digraph. By this assumption, it is proved that all the players are able to exchange and update all the estimates of the actions interfering with their cost functions.
	
	The paper is organized as follows. In Section II, the problem
	statement and assumptions are provided for the game with a complete interference digraph. An asynchronous gossip-based
	algorithm is proposed in Section III. In Section IV, convergence
	of the algorithm with diminishing step sizes is discussed. In Sections V and VI, the problem statement and the proposed algorithm for the game with a partial interference digraph are investigated, respectively and its convergence to an NE of the game is proved in Section VII. Simulation results are
	presented in Section VIII and concluding remarks are provided in Section IX.
	\section{Problem Statement: Game With a Complete Interference Digraph}\label{problem_statement}
	
	Consider a multi-player game in a network with a set of players $V$. The interference of players' actions on the cost functions is represented by a complete \emph{interference digraph} $G(V,E)$, with $E$ marking the pair of players that interfere one with another. Note that for a complete digraph every pair of distinct nodes is connected by a pair of unique edges (one in each direction). 
	
	The game is denoted by $\mathcal{G}(V,\Omega_i,J_i)$ and defined over
	\begin{itemize}
		\item $V=\{1,\ldots,N\}$: Set of players,
		\item $\Omega_i\subset\mathbb{R}$: Action set of player $i$, $\forall i\in V$ with $\Omega=\prod_{i\in V}\Omega_i\subset\mathbb{R}^N$ the action set of all players,
		\item $J_i:\Omega\rightarrow \mathbb{R}$: Cost function of player $i$, $\forall i\in V$,
	\end{itemize}
	In the following we define a few notations for players' actions.
	\begin{itemize}
		\item $x=(x_i,x_{-i})\in\Omega$: All players actions,
		\item $x_i\in\Omega_i$: Player $i$'s action, $\forall i\in V$,
		\item $x_{-i}\in\Omega_{-i}:=\prod_{j\in V\backslash\{i\}}\Omega_j$: All other players' actions except $i$.
	\end{itemize} 
	The game is defined as a set of $N$ simultaneous optimization problems as follows:
	\begin{equation}
	\label{mini_0}
	\begin{cases}
	\underset{y_i}{\text{minimize}}& J_i(y_i,x_{-i}) \\
	\text{subject to}& y_i\in \Omega_i
	\end{cases}\quad\forall i\in V.
	\end{equation}
	Each problem is run by an individual player and its solution is dependent on the solution of the other problems.
	
	The objective is to find an NE of this game which is defined as follows:
	\begin{definition}\label{Nash_def}
		Consider an $N$-player game $\mathcal{G}(V,\Omega_i,J_i)$, each player $i$ minimizing the cost function $J_i:\Omega\rightarrow\mathbb{R}$. A vector $x^*=(x_i^*,x_{-i}^*)\in\Omega$ is called an NE of this game if
		\begin{equation}
		J_i(x_i^*,{x_{-i}^{*}})\leq J_i(x_{i},{x_{-i}^{*}})\quad\forall x_i\in \Omega_i,\,\,\forall i\in V.
		\end{equation}
	\end{definition}
	Note that for game \eqref{mini_0}, the NE lies in the intersection of the solutions of the optimization problems. We state a few assumptions for the existence and the uniqueness of an NE.
	\begin{assumption}
		\label{assump}
		For every $i\in V$, 
		\begin{itemize}
			\item $\Omega_i$ is non-empty, compact and convex, 
			\item $J_i(x_i,x_{-i})$ is $C^1$ in $x_i$ and continuous in $x$, \item $J_i(x_i,x_{-i})$ is convex in $x_i$ for every $x_{-i}$.
		\end{itemize}
	\end{assumption}
	\hspace{-0.4cm}The compactness of $\Omega$ implies that $\forall i\in V$ and $x\in\Omega$, 
	\begin{equation}\label{bounded}
	\|\nabla_{x_i}J_i(x)\|\leq C,\quad\text{for some }C>0.
	\end{equation}
	Let $F:\Omega\rightarrow\mathbb{R}^N$, $F(x):=[\nabla_{x_i}J_i(x)]_{i\in V}$ be the pseudo-gradient vector of the cost functions (game map).
	\begin{assumption}\label{Lip_assump}
		$F$ is strictly monotone,
		\begin{equation*}
		(F(x)-F(y))^T(x-y)> 0\quad\forall x,y\in \Omega,\text{ }x\neq y.
		\end{equation*}
	\end{assumption}
	\begin{assumption}\label{Lip_assump2}
		$\nabla_{x_i}J_i(x_i,u)$ is Lipschitz continuous in $x_i$, for every fixed $u\in\Omega_{-i}$ and for every $i\in V$, i.e., there exists $\sigma_i>0$ such that
		\begin{equation*}
		\|\nabla_{x_i}J_i(x_i,u)-\nabla_{x_i}J_i(y_i,u)\|\leq \sigma_i\|x_i-y_i\|\quad\forall x_i,y_i\in\Omega_{i}.
		\end{equation*}
		Moreover, $\nabla_{x_i}J_i(x_i,u)$ is Lipschitz continuous in $u$ with a Lipschitz constant $L_i>0$ for every fixed $x_i\in\Omega_i,\,\forall i\in V$.
	\end{assumption}
	\begin{remark}\label{Lipscitz_F}
		Assumption~\ref{Lip_assump2} implies that $\nabla_{x_i}J_i(x)$ and $F(x)$ are Lipschitz continuous in $x\in\Omega$ with Lipschitz constants $\rho_i=\sqrt{2L_i^2+2\sigma_i^2}$ and $\rho=\sqrt{\sum_{i\in V}\rho_i^2}$, respectively for every $i\in V$.
	\end{remark}
	
	In game \eqref{mini_0}, the only information available to each player $i$ is $J_i$ and $\Omega$. Thus, each player maintains an estimate of the other players actions and exchanges those estimates with the local neighbors to update them.
	A \emph{communication digraph} $G_C(V,E_C)$ is defined where $E_C\subseteq V\times V$ denotes the set of communication links between the players. $(i,j)\in E_C$ if and only if player $i$ sends his information to player $j$. Note that $(i,j)\in E_C$ does not necessarily imply $(j,i)\in E_C$. The set of in-neighbors of player $i$ in $G_C$, denoted by $N_C^\text{in}(i)$, is defined as $N_C^\text{in}(i):=\{j\in V|(j,i)\in E_C\}$. The following assumption on $G_C$ is used. 
	\begin{assumption}\label{connectivity}
		$G_C$ is strongly connected.
	\end{assumption}
	A digraph is called strongly connected if there exists a path between each ordered pair of vertices of the digraph.
	
	Our objective is to find an algorithm for computing an NE of $\mathcal{G}(V,\Omega_i,J_i)$ using only imperfect information over the communication digraph $G_C(V,E_C)$.
	\section{Asynchronous Gossip-based Algorithm}\label{asynch}
	We propose a projected gradient-based algorithm using an asynchronous gossip-based method as in \cite{salehisadaghiani2016distributed}. The algorithm can be briefly explained as follows:
	\begin{itemize}
		\item Each player maintains a \emph{temporary estimate} of all players' actions. 
		\item They receive information from the local neighbors over $G_C$ to update their temporary estimates.
		\item Then, they solve their own optimization problems using their \emph{final estimates} and update their actions.
	\end{itemize}
	The algorithm is inspired by \cite{salehisadaghiani2016distributed} except that the communications are supposed to be directed in a sense that the information exchange is considered over a directed path. Our challenge here is to deal with the asymmetric communications between the players. This makes the convergence proof dependent on a {\emph{non-doubly stochastic weight matrix}}, whose properties need to be investigated and proved. In many cases e.g. sensor networks, symmetric communication may not be of interest and could be an undesirable feature.
	
	The algorithm is elaborated as follows:\\	
	1- \textbf{\emph{Initialization Step:}}
	Each player $i$ maintains an initial \emph{temporary} estimate $\tilde{x}^i(0)\in\Omega$ for all players. Let $\tilde{x}_j^i(0)\in\Omega_j\subset\mathbb{R}$ be player $i$'s initial temporary estimate of player $j$'s action, for $i,j\in V$.\\
	2- \textbf{\emph{Gossiping Step:}}
	At iteration $k$, player $i_k$ becomes active uniformly at random and selects a communication in-neighbor indexed by $j_k\in N_C^\text{in}(i_k)$ uniformly at random. Let $\tilde{x}^i(k)\in\Omega\subset\mathbb{R}^N$ be player $i$'s temporary estimate at iteration $k$. Then player $j_k$ sends his temporary estimate $\tilde{x}^{j_k}(k)$ to player $i_k$.
	After receiving the information, player $i_k$ constructs his final estimate of all players. Let $\hat{x}_j^{i}(k)\in\Omega_j\subset\mathbb{R}$ be player $i$'s final estimate of player $j$'s action, for $i,j\in V$.
	The final estimates are computed as in the following:
	\begin{enumerate}
		\item Players $i_k$'s final estimate:
		\begin{equation}\label{excluding}
		\hspace{-0.75cm}\begin{cases}
		\hat{x}_{i_k}^{i_k}(k)=\tilde{x}_{i_k}^{i_k}(k)\\
		\hat{x}_{-i_k}^{i_k}(k)=\frac{\tilde{x}_{-i_k}^{i_k}(k)+\tilde{x}_{-i_k}^{j_k}(k)}{2}.
		\end{cases}
		\end{equation}
		Note that  $\tilde{x}_i^i(k)=x_i(k)$ for all $i\in V$, since no estimate is needed for the players' own actions.\\ 
		\item For all other players $i\neq i_k$, the temporary estimate is maintained, i.e.,
		\begin{equation}\label{other_inc_exc}
		\hat{x}^i(k)=\tilde{x}^i(k),\quad\forall i\neq i_k.
		\end{equation}
	\end{enumerate}
	
	We use communication weight matrix $W(k):=[w_{ij}(k)]_{i,j\in V}$ to obtain a compact form of the gossip protocol. $W(k)$ is {\emph{a non-doubly (row) stochastic weight matrix}} defined as follows:
	\begin{equation}
	\label{weight_matrix}
	W(k)=I_N-\frac{e_{i_k}(e_{i_k}-e_{j_k})^T}{2},
	\end{equation}
	where $e_i\in\mathbb{R}^N$ is a unit vector. Note that $W(k)$ is different from the one used in \cite{salehisadaghiani2016distributed}. The non-doubly (row) stochasticity of $W(k)$ is translated into:
	\begin{equation}
	W(k)\mathbf{1}_N=\mathbf{1}_N,\qquad\mathbf{1}_N^TW(k)\neq\mathbf{1}_N^T.\label{non-doubly2}
	\end{equation}
	
	Let $\bar{x}(k)=[\bar{x}^1(k),\ldots,\bar{x}^N(k)]^T\in\Omega^N$ be an intermediary variable such that
	\begin{equation}
	\label{dummy}
	\bar{x}(k)=(W(k)\otimes I_N)\tilde{x}(k),
	\end{equation}
	where $\tilde{x}(k)=[\tilde{x}^1(k),\ldots,\tilde{x}^N(k)]^T\in \Omega^N$ is the overall temporary estimate at $k$. Using \eqref{weight_matrix} one can combine \eqref{excluding} and \eqref{other_inc_exc} in a compact form of $\hat{x}_{-i_k}^{i_k}(k)=\bar{x}_{-i_k}^{i_k}(k)$ and $\hat{x}^i(k)=\bar{x}^i(k)$ for $\forall i\neq i_k$.\\
	3- \textbf{\emph{Local Step}}
	
	At this moment all the players update their actions according to a projected gradient-based method. Let $\bar{x}^i=(\bar{x}_i^i,\bar{x}_{-i}^i)\in\Omega$, $\forall i\in V$ with $\bar{x}_i^i\in\Omega_i$ be the intermediary variable associated to player $i$. Because of imperfect information available to player $i$, he uses $\bar{x}_{-i}^i(k)$  and updates his action as follows: if $i=i_k$,
	\begin{equation}\label{local_step}
	x_i(k+1)=T_{\Omega_i}[x_i(k)-\alpha_{k,i} \nabla_{x_i}J_i(x_i(k),\bar{x}_{-i}^i(k))],
	\end{equation}
	otherwise, $x_i(k+1)=x_i(k)$. In \eqref{local_step}, $T_{\Omega_i}:\mathbb{R}\rightarrow\Omega_i$ is an Euclidean projection and $\alpha_{k,i}$ are diminishing step sizes such that
	\begin{equation}
	\label{diminish_step}
	\sum_{k=1}^{\infty}\alpha_{k,i}^2<\infty,\qquad\sum_{k=1}^{\infty}\alpha_{k,i}=\infty\quad \forall i\in V.
	\end{equation}
	Note that $\alpha_{k,i}$ is inversely related to the number of updates $\nu_k(i)$ that each player $i$ has made until time $k$ (i.e., $\alpha_{k,i}=\frac{1}{\nu_k(i)}$). In \eqref{local_step}, the players who are not involved in communication at iteration $k$ maintain their actions unchanged.
	At this moment the updated actions are available for players to update their temporary estimates for every $i\in V$ as follows:
	\begin{equation}\label{temp_update}
	\tilde{x}^i(k+1)=\bar{x}^i(k)+(x_i(k+1)-\bar{x}_i^i(k))e_i,\quad\forall i\in V.\end{equation}
	At this point, the players are ready to begin a new iteration from step 2.
	\begin{algorithm}
		\caption{}
		\begin{algorithmic}[1]
			\State \textbf{initialization} $\tilde{x}^i(0)\in\Omega\quad\forall i\in V$
			\For{$k=1,2,\ldots$ }
			\State \hspace{-0.5cm}$i_k\in V$ and $j_k\in N_C^\text{in}(i_k)$ communicate.
			\State \hspace{-0.5cm}$W(k)=I_N-\frac{e_{i_k}(e_{i_k}-e_{j_k})^T}{2}$.
			\State \hspace{-0.5cm}$\bar{x}(k)=(W(k)\otimes I_N)\tilde{x}(k)$.
			\State \hspace{-0.5cm}$x_{i_k}(k\!+\!1)\!=\!T_{\Omega_{i_k}}[x_{i_k}(k)\!-\!\alpha_{k,{i_k}} \nabla_{x_i}J_{i_k}(x_{i_k}(k)\!,\!\bar{x}_{-{i_k}}^{i_k}(k))]$,
			\Statex	$x_i(k+1)=x_i(k)$, if $i\neq i_k$.
			\State \hspace{-0.5cm}$\tilde{x}^i(k+1)=\bar{x}^i(k)+(x_i(k+1)-\bar{x}_i^i(k))e_i,\quad\forall i\in V$.
			\EndFor
		\end{algorithmic}
	\end{algorithm}
	
	Now we elaborate on the non-doubly stochastic property of $W(k)$ from two perspectives. 
	\begin{enumerate}
		\item \textbf{Design}: By the row (non-doubly) stochastic property of $W(k)$, the temporary estimates remain at consensus subspace once they reach there. This can be verified by \eqref{dummy} when $\tilde{x}(k)=\textbf{1}_N\otimes\vec{\alpha}$ for an $N\times1$ vector $\vec{\alpha}$, since,
		\begin{equation}
		\label{dummy11}
		\bar{x}(k)=(W(k)\otimes I_N)(\textbf{1}_N\otimes\vec{\alpha})=\textbf{1}_N\otimes\vec{\alpha}.
		\end{equation} 
		Equation \eqref{dummy11} along with \eqref{local_step} and \eqref{temp_update} imply that the consensus is maintained. 
		On the other hand {\emph{$W(k)$ is not a column-stochastic matrix}} which is a critical property used in \cite{salehisadaghiani2016distributed}. This implies that the average of temporary estimates is not equal to the average of $\bar{x}$. Indeed by \eqref{dummy}, 
		\begin{eqnarray}
		\label{dummy111}
		\hspace{-0.5cm}\frac{1}{N}(\textbf{1}_N^T\otimes I_N)\bar{x}(k)=\frac{1}{N}(\textbf{1}_N^T\otimes I_N)(W(k)\otimes I_N)\tilde{x}(k)\neq\frac{1}{N}(\textbf{1}_N^T\otimes I_N)\tilde{x}(k).
		\end{eqnarray}
		{\emph{Equation~\eqref{dummy111} along with \eqref{local_step} and \eqref{temp_update} imply that the average of temporary estimates is not preserved for the next iteration. Thus, it seems infeasible to obtain an exact convergence toward the average consensus \cite{aysal2009broadcast}}}. Instead we show an almost sure (a.s.) convergence\footnote[1]{Almost sure convergence: Random variables converge with probability~1.} of the temporary estimates toward an average consensus\footnote[2]{The same objective is followed by \cite{nedic2011asynchronous} to find a broadcast gossip algorithm (with non-doubly stochastic weight matrix) in the area of distributed optimization. However, in the proof of Lemma~2 (\cite{nedic2011asynchronous} page~1348) which is mainly dedicated to this discussion, the \underline{doubly stochasticity} of $W(k)$ is used right after equation (22) which violates the main assumption on $W(k)$.}.
		\item \textbf{Convergence Proof}: $\lambda_{\max}(W(k)^TW(k))$ is a key parameter in the convergence proof (as in \cite{nedic2011asynchronous,salehisadaghiani2016distributed}). Unlike \cite{salehisadaghiani2016distributed}, the non-doubly stochastic property of $W(k)^TW(k)$ ends up in having $\lambda_{\max}(W(k)^TW(k))>1$. We resolve this issue in Lemma~\ref{lemma_nondoubly}.
	\end{enumerate}
	\section{Convergence For Diminishing Step Sizes}\label{convergence_diminish}
	In this section we prove convergence of the algorithm for diminishing step sizes as in \eqref{diminish_step}. 
	
	Consider a memory in which the history of the decision making is recorded. Let $\mathcal{M}_k$ denote the {\emph{sigma-field}} generated by the history up to time $k-1$ with $\mathcal{M}_0=\{\tilde{x}^i(0),\text{ }i\in V\}$.
	\begin{equation}\mathcal{M}_k=\mathcal{M}_0\cup\Big\{(i_l,j_l); 1\leq l\leq k-1\Big\},\quad \forall k\geq 2.\end{equation}
	In the proof we use a well-known result on super martingale convergence, (Lemma~11, Chapter~2.2,  \cite{polyak1987introduction}).
	\begin{lemma}
		\label{sigmond}
		\label{dge2}
		Let $V_k$, $u_k$, $\beta_k$ and $\zeta_k$ be non-negative random variables adapted to $\sigma$-algebra $\mathcal{M}_k$. If $\sum_{k=0}^{\infty}u_k<\infty$ a.s., $\sum_{k=0}^{\infty}\beta_k<\infty$ a.s., and $\mathbb{E}[V_{k+1}|\mathcal{M}_k]\leq(1+u_k)V_k-\zeta_k+\beta_k$ a.s. for all $k\geq 0$, then $V_k$ converges a.s. and $\sum_{k=0}^{\infty}\zeta_k<\infty$ a.s.
	\end{lemma}
	As explained in the design challenge in Section~\ref{asynch}, we consider a.s. convergence.
	Convergence is shown in two parts. First, we prove a.s. convergence of the temporary estimate vectors $\tilde{x}^i,$ to an average consensus, proved to be the vectors' average. Then we prove a.s. convergence of players' actions toward an NE.
	
	Let $\tilde{x}(k)$ be the overall temporary estimate vector. The average of all temporary estimates at $T(k)$ is defined as follows:
	\begin{equation}\label{average}
	Z(k)=\frac{1}{N}(\textbf{1}_N^T\otimes I_N)\tilde{x}(k).
	\end{equation}
	As mentioned in Section~\ref{asynch}, the major difference between the proposed algorithm and the one in \cite{salehisadaghiani2016distributed} is in using a non-doubly stochastic weight matrix $W(k)$ which was a key step. The following lemma is used to overcome these challenges.
	\begin{lemma}\label{lemma_nondoubly}
		Let $Q(k)=(W(k)-\frac{1}{N}\textbf{1}_N\textbf{1}_N^TW(k))\otimes I_N$ and $W(k)$ be a non-doubly (row) stochastic weight matrix defined in \eqref{weight_matrix} which satisfies \eqref{non-doubly2}. Let also $\gamma=\lambda_{\max}\big(\mathbb{E}[Q(k)^TQ(k)]\big)$. Then $\gamma<1$.
	\end{lemma}
	\par{\emph{Proof}}.
	Consider the variational characterization of $\gamma$. Since $\mathbb{E}[Q(k)^TQ(k)]$ is an $N^2\times N^2$ symmetric matrix, we can write,
	\begin{equation}\label{gamma_optimiz}
	\gamma=\sup_{x\in\mathbb{R}^{N^2},\|x\|=1}x^T\mathbb{E}[Q(k)^TQ(k)]x\geq 0.
	\end{equation}
	Due to space limitation we drop the constraints of $\text{sup}(\cdot)$. By the definition of $Q(k)$, we obtain,
	\begin{eqnarray}
	\gamma\!=\!\sup_{x} x^T\mathbb{E}\Big[\!\Big(\!W(k)^T\!W(k)\!-\!\frac{1}{N}W(k)^T\!\textbf{1}_N\!\textbf{1}_N^TW(k)\!\Big)\!\otimes\! I_N\!\Big]x.\nonumber
	\end{eqnarray}
	Using \eqref{weight_matrix}, we expand $\gamma$ as follows:
	\begin{eqnarray}\label{term1-term2}
	&&\gamma=\sup_{x} x^T\mathbb{E}\Big[\Big\{\Big(\underbrace{I_N-\frac{1}{4N}(e_{i_k}-e_{j_k})(e_{i_k}-e_{j_k})^T}_{\text{Term}~1}\Big)\nonumber\\
	&&-\Big(\underbrace{\frac{1}{N}\textbf{1}_N\textbf{1}_N^T+\frac{1}{2}(e_{i_k}-\frac{1}{N}\textbf{1}_N)(e_{i_k}-e_{j_k})^T+\frac{1}{2}(e_{i_k}-e_{j_k})(e_{i_k}-\frac{1}{N}\textbf{1}_N)^T}_{\text{Term}~2}\nonumber\\
	&&\underbrace{-\frac{1}{4}(e_{i_k}-e_{j_k})(e_{i_k}-e_{j_k})^T}_{\text{Term}~2}\Big)\Big\}\otimes I_N\Big]x.
	\end{eqnarray}
	Note that $\mathbb{E}[(\text{Term}~1-\text{Term}~2)\otimes I_N]$ is a symmetric matrix.\\
	\textbf{Claim~1}: For all  $x\in\mathbb{R}^{N^2},\|x\|=1$, we have, $x^T\mathbb{E}\Big[\text{Term~1}\otimes I_N\Big]x\leq1$.
	The equality only holds for $x=\textbf{1}_N\otimes y$ where $y\in\mathbb{R}^{N}$ and $\|y\|=\frac{1}{\sqrt{N}}$.\\
	\textbf{Proof of Claim~1}: Multiplying $x^T$ and $x$ into the argument of the expected value in \eqref{term1-term2} and using $\|x\|=1$, we obtain,
	\begin{eqnarray}\label{term1esbat}
	x^T\mathbb{E}\Big[\text{Term~1}\otimes I_N\Big]x=1-\frac{1}{4N}\mathbb{E}\Big[\Big\|\Big((e_{i_k}-e_{j_k})^T\otimes I_N\Big)x\Big\|^2\Big]\leq1.
	\end{eqnarray}
	The equality holds only when,
	\begin{eqnarray}\label{kolle_ivajha}
	&&\mathbb{E}\Big[\Big\|\Big((e_{i_k}-e_{j_k})^T\otimes I_N\Big)x\Big\|^2\Big]=0\Leftrightarrow\Big\|\Big((e_{i_k}-e_{j_k})^T\otimes I_N\Big)x\Big\|^2=0
	\Leftrightarrow\nonumber\\
	&&(e_{i_k}^T\otimes I_N)x=(e_{j_k}^T\otimes I_N)x.
	\end{eqnarray}
	Equation \eqref{kolle_ivajha} holds for all $k\geq0$, $i_k\in V$ and $j_k\in N_C^\text{in}(i_k)$. By the strong connectivity of $G_C$ (Assumption~\ref{connectivity}), \eqref{kolle_ivajha} becomes $(e_{i}^T\otimes I_N)x=(e_{j}^T\otimes I_N)x,\quad\forall i,j\in V$ which implies that $x=\textbf{1}_N\otimes y$ where $y\in\mathbb{R}^{N}$. Moreover, $\|x\|=1$ yields,
	\begin{eqnarray}
	\|\textbf{1}_N\otimes y\|^2=1\Leftrightarrow(\textbf{1}_N^T\otimes y^T)(\textbf{1}_N\otimes y)=1\Leftrightarrow\|y\|=\frac{1}{\sqrt{N}}.\nonumber
	\end{eqnarray}
	\textbf{Claim~2}: For  $x=(\textbf{1}_N\otimes y)\in\mathbb{R}^{N^2}$ where $y\in\mathbb{R}^{N}$ and $\|y\|=\frac{1}{\sqrt{N}}$ we have $x^T\mathbb{E}\Big[\text{Term~2}\otimes I_N\Big]x>0$.\\
	\textbf{Proof of Claim~2}: For  $x=\textbf{1}_N\otimes y$ and $\|y\|=\frac{1}{\sqrt{N}}$ we obtain by the mixed product property of Kronecker that,
	\begin{eqnarray}\label{term2mosavi}
	\hspace{-0.7cm}&&x^T\mathbb{E}\Big[\!\text{Term~2}\!\otimes\! I_N\!\Big]\!x\!=\!\mathbb{E}\Big[\!(\textbf{1}_N^T\!\otimes\! y^T)(\text{Term~2}\otimes I_N)(\textbf{1}_N\otimes y)\Big]\nonumber\\
	\hspace{-0.7cm}&&=\mathbb{E}\Big[\Big(\textbf{1}_N^T(\text{Term~2})\textbf{1}_N\Big)\otimes y^Ty\Big].
	\end{eqnarray} 
	It is straightforward to verify that $\textbf{1}_N^T(\text{Term~2})\textbf{1}_N=N$ because all the summands in Term~2 except the first one vanish by multiplying $\textbf{1}_N^T$ and $\textbf{1}_N$. Having that $y^Ty=\frac{1}{N}$, \eqref{term2mosavi} implies $x^T\mathbb{E}\Big[\text{Term~2}\otimes I_N\Big]x=1>0$.
	By Claims~1,~2 and using the fact that Terms~1,~2 are symmetric and $\gamma\geq 0$, \eqref{term1-term2} implies that $\gamma<1$.
	$\hfill\blacksquare$ 
	
	We use Lemmas~\ref{sigmond}, \ref{lemma_nondoubly} to show $\tilde{x}(k)$ converges a.s. to $Z(k)$.
	\begin{theorem}\label{consensus1}
		Let $\tilde{x}(k)$ be the stack vector with all temporary estimates of the players and $Z(k)$ be its average as in \eqref{average}. Let also $\alpha_{k,\text{max}}\!=\!\max_{i\in V}\!\alpha_{k,i}$. Then under Assumptions~\ref{assump},\ref{connectivity}, the following hold.
		\begin{enumerate}[i)]
			\item $\sum_{k=0}^{\infty}\alpha_{k,\text{max}}\|\tilde{x}(k)-(\textbf{1}_N\otimes I_N)Z(k)\|<\infty$ a.s.,
			\item $\sum_{k=0}^{\infty}\|\tilde{x}(k)-(\textbf{1}_N\otimes I_N)Z(k)\|^2<\infty$ a.s.
		\end{enumerate}
	\end{theorem}
	\par{\emph{Proof}}. The proof follows as in the proof of Theorem~1 in \cite{salehisadaghiani2016distributed}, but the critical step here is in using Lemma~\ref{lemma_nondoubly}.

	Theorem~\ref{consensus1} yields the following corollary for $x(k)$ and $\bar{x}(k)$.
	\begin{corollary}\label{remark}
		For the players' actions $x(k)$ and $\bar{x}(k)$, the following terms hold a.s. under Assumptions~\ref{connectivity}-\ref{assump}.\\
		i) $\sum_{k=0}^{\infty}\alpha_{k,\text{max}}\|x(k)-Z(k)\|<\infty$ a.s.,\qquad ii) $\sum_{k=0}^{\infty}\|x(k)-Z(k)\|^2<\infty$ a.s.,\\
		iii) $\sum_{k=0}^{\infty}\mathbb{E}\Big[\|\bar{x}(k)-(\textbf{1}_N\otimes I_N)Z(k)\|^2\Big |\mathcal{M}_{k}\Big]<\infty$ a.s.
	\end{corollary}
	\par{\emph{Proof}}.
	The proof follows directly from Theorem~\ref{consensus1} noting that $x(k)=[\tilde{x}_i^i(k)]_{i\in V}$ and $\bar{x}(k)=(W(k)\otimes I_N)\tilde{x}(k)$ \eqref{dummy}.
	\begin{theorem}
		\label{Prop_algo_1}
		Let $x(k)$ and $x^*$ be the players' actions and the NE of $\mathcal{G}$, respectively. Under Assumptions~\ref{connectivity}-\ref{Lip_assump2}, the sequence $\{x(k)\}$ generated by the algorithm converges to $x^*$, almost surely.
	\end{theorem}
	\vspace{-0.2cm}
	\par{\emph{Proof}}. The proof is similar to the proof of Theorem~2 in \cite{salehisadaghiani2016distributed} based on Theorem~\ref{consensus1}. 

	Theorem~\ref{Prop_algo_1} verifies that the actions of the players converge a.s. toward the NE using the fact that the actions converge to a consensus subspace (Corollary~\ref{remark}).
	\section{Game With a Partial Interference Digraph}\label{Section_Interference}
	We extend the game defined in Section~\ref{problem_statement} to the case with partially coupled cost functions, in the sense that the cost functions may be affected by the actions of any subset of players. The game is denoted by $\mathcal{G}(V,G_I,\Omega_i,J_i)$ where $G_I(V,E_I)$ is an interference digraph with $E_I$ marking the players whose actions interfere the other players' cost functions, e.g.,
	\begin{figure}[H]
		\vspace{-1.5cm}
		\hspace{-9cm}
		\centering
		\includegraphics [scale=0.7]{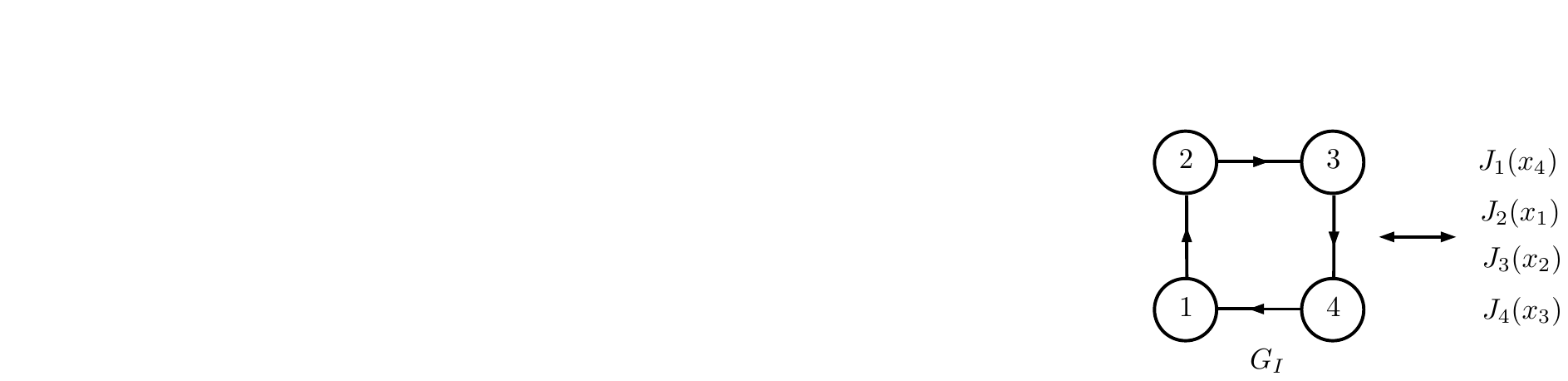}
		\vspace{-0.4cm}
	\end{figure}
	We also denote by $N_I^\text{in}(i):=\{j\in V|(j,i)\in E_I\}$, the set of in-neighbors of player $i$ in $G_I$ whose actions affect $J_i$ and $\tilde{N}_I^\text{in}(i):=N_I^\text{in}(i)\cup\{i\}$.
	
	The following assumption is considered for $G_I$.
	\begin{assumption}\label{G_I_Strongly_Connected}
		$G_I$ is strongly connected.
	\end{assumption}
	The cost function of player $i$, $J_i$, $\forall i\in V$, is defined over $\Omega^i\rightarrow \mathbb{R}$ where $\Omega^i=\prod_{j\in\tilde{N}_I^\text{in}(i)}\Omega_j\subset\mathbb{R}^{|\tilde{N}_I^\text{in}(i)|}$ is the action set of players interfering with the cost function of player $i$.
	
	A few notations for players' actions are given:
	\begin{itemize}
		\item $x^i=(x_i,x_{-i}^i)\in\Omega^i$: All players' actions which interfere with $J_i$,
		\item $x_{-i}^i\in\Omega_{-i}^i:=\prod_{j\in N_I^\text{in}(i)}\Omega_j$: Other players' actions which interfere with $J_i$.
	\end{itemize}
	Given $x_{-i}^i$, each player $i$ aims to minimize his own cost function selfishly,
	\begin{equation}
	\label{mini_0_interference}
	\begin{cases}
	\underset{y_i}{\text{minimize}}& J_i(y_i,x_{-i}^i) \\
	\text{subject to}& y_i\in \Omega_i
	\end{cases}\quad\forall i\in V.
	\end{equation}
	
	Known parameters to player $i$ are as follows: 1) Cost function of player $i$, $J_i$ and 2) Action set $\Omega^i$. Note that this game setup is similar to the one in \cite{ssalehisadaghiani2016distributed} except for a directed $G_C$ used for asymmetric communications.
	
	A restatement of an NE definition (Definition~\ref{Nash_def}) adapted to the interference digraph $G_I$ is defined as follows:
	\begin{definition}\label{Nash_def_Jadid}
		Consider an $N$-player game $\mathcal{G}(V,G_I,\Omega_i,J_i)$, each player $i$ minimizing the cost function $J_i:\Omega^i\rightarrow\mathbb{R}$. A vector $x^*=(x_i^*,x_{-i}^*)\in\Omega$ is called an NE of this game if for every given $x_{-i}^{i*}\in\Omega_{-i}^i$
		\begin{equation}
		J_i(x_i^*,{x_{-i}^{i*}})\leq J_i(x_{i},{x_{-i}^{i*}})\quad\forall x_i\in \Omega_i,\,\,\forall i\in V.
		\end{equation}
	\end{definition}
	
	Our first objective is to design an assumption on $G_C$ such that all required information is communicated by the players after sufficiently many iterations. In other words, we ensure that player $i$, $\forall i\in V$ receives information from all the players whose actions interfere with his cost function.
	\begin{definition}\label{transitive_reduction}
		{Transitive reduction}: A digraph $H$ is a transitive reduction of $G$ which is obtained as follows: for all three vertices $i,j,l$ in $G$ such that edges $(i,j)$, $(j,l)$ are in $G$, $(i,l)$ is removed from $G$.
	\end{definition} 
	In simple terms, a transitive reduction of a digraph is a digraph without the parallel paths between the vertices.
	\begin{remark}
		Transitive reduction is different from the notion of {\emph{maximal triangle-free spanning subgraph}}, which is used in Assumption~2 in \cite{ssalehisadaghiani2016distributed}.
	\end{remark}
	\begin{assumption}\label{connectivity_GTR}
		The following holds for the communication graph $G_C$:
		\begin{itemize}
			\item $G_\text{TR}\subseteq G_C\subseteq G_I$, where $G_\text{TR}$ is a transitive reduction of $G_I$.
		\end{itemize}
	\end{assumption}
	\begin{remark}\label{G_C_Strong}
		$G_C$ is strongly connected because it is a superset of the transitive reduction of $G_I$. Note that the transitive reduction preserves the strong connectivity of a digraph by removing only the parallel paths.
	\end{remark}	
	\begin{lemma}\label{TR_lemma}
		Let $G_I$ and $G_C$ satisfying Assumptions~\ref{G_I_Strongly_Connected}, \ref{connectivity_GTR}. Then,$\forall i\in V$,
		\begin{equation}\label{Lemma_set}
		\bigcup_{j\in N^{\text{in}}_C(i)}\big(N^{\text{in}}_I(i)\cap\tilde{N}^{\text{in}}_I(j)\big)=N^{\text{in}}_I(i).
		\end{equation}
	\end{lemma}
	\par{\emph{Proof}}. The proof is similar to the proof of Lemma~2 in \cite{salehisadaghiani2017adistribute}, but modified to adapt for the directed graph. We need to show $N^{\text{in}}_I(i)\subseteq\bigcup_{j\in N^{\text{in}}_C(i)}\tilde{N}^{\text{in}}_I(j)$ $\forall i\in V$ from which it is straightforward to deduce \eqref{Lemma_set}.
	
	For the case when $G_C=G_I$, we obtain,
	\begin{equation}\label{Lemma_G_C=G_I}
	\bigcup_{j\in N^{\text{in}}_C(i)}\!\tilde{N}^{\text{in}}_I(j)\!=\!\bigcup_{j\in N^{\text{in}}_I(i)}\!\tilde{N}^{\text{in}}_I(j)\!\supseteq\!\bigcup_{j\in {N}^{\text{in}}_I(i)}\!\{j\}\!=\!N^{\text{in}}_I(i).
	\end{equation}
	In \eqref{Lemma_G_C=G_I}, we used $\{j\}\subseteq\tilde{N}^{\text{in}}_I(j)$ by the definition of $\tilde{N}^{\text{in}}_I(j)$.
	
	Now assume that $G_\text{TR}\subseteq G_C\subset G_I$. To prove \eqref{Lemma_set}, it is sufficient to show that $N^{\text{in}}_I(i)\subseteq\bigcup_{j\in N^{\text{in}}_\text{TR}(i)}\tilde{N}^{\text{in}}_\text{TR}(j)$, where $N^{\text{in}}_\text{TR}(i)$ is the set of in-neighbors of player $i$ in $G_\text{TR}$ and $\tilde{N}^{\text{in}}_\text{TR}(i)$ in addition to $N^{\text{in}}_\text{TR}(i)$ contains $\{i\}$. In other words we need to show that any in-neighbor of player $i$ (any vertex with an incoming edge to $i$) in $G_I$ is either an in-neighbor or \enquote{in-neighbor of an in-neighbor} of player $i$ (a vertex with an incoming path of at most length 2 to $i$) in $G_\text{TR}$. If an incoming edge to $i$ exists both in $G_I$ and $G_\text{TR}$, the corresponding in-neighbor of $i$ in $G_I$ is an in-neighbor of $i$ in $G_\text{TR}$. Otherwise, if there exists an incoming edge to $i$ in $G_I$ that is missing in $G_\text{TR}$, according to Definition~\ref{transitive_reduction}, there exists a directed path of length 2 parallel to the missing edge in $G_{\text{TR}}$. So the corresponding in-neighbor of player $i$ in $G_I$ is an in-neighbor of an in-neighbor of player $i$ in $G_\text{TR}$.
	$\hfill\blacksquare$
	\begin{remark}
		Lemma~\ref{TR_lemma} verifies that if $G_I$ and $G_C$ satisfy Assumptions~\ref{G_I_Strongly_Connected}, \ref{connectivity_GTR}, then player $i\in V$ exchanges all of the estimates of the players' actions which interfere with his cost function after enough number of communications (see equation \eqref{excluding}).
	\end{remark}
	The assumptions for existence and uniqueness of an NE are Assumptions~\ref{assump}-\ref{Lip_assump2} with the cost functions adapeted to $G_I$.
	
	Our second objective is to find an algorithm for computing an NE of $\mathcal{G}(V,G_I,\Omega_i,J_i)$ with partially coupled cost functions as described by the directed graph $G_I(V,E_I)$ using only imperfect information over $G_C(V,E_C)$.\vspace{-0cm}
	\section{Asynchronous Gossip-based Algorithm adapted to $G_I$}\label{asynch_interference}
	The structure of the algorithm is similar to the one in Section~\ref{asynch}. The steps are elaborated in the following:\\
	1- \textbf{\emph{Initialization Step:}}
	\begin{itemize}
		\item $\tilde{x}^i(0)\in\Omega^i$: Player $i$'s initial temporary estimate.
	\end{itemize}
	2- \textbf{\emph{Gossiping Step:}}
	\begin{itemize}
		\item $\tilde{x}_j^i(k)\in\Omega_j\subset\mathbb{R}$: Player $i$'s temporary estimate of player $j$'s action at iteration $k$.
		\item $\hat{x}_j^{i}(k)\in\Omega_j\subset\mathbb{R}$: Player $i$'s final estimate of player $j$'s action at iteration $k$, for $i\in V,\,j\in\tilde{N}_I^\text{in}(i)$.
		\item Final estimate construction:
		\begin{eqnarray}\label{gossip_interference}
		\hspace{-1cm}&&\hat{x}_{l}^{i_k}(k)=\begin{cases}
		\frac{\tilde{x}_{l}^{i_k}(k)+\tilde{x}_{l}^{j_k}(k)}{2},& l\in (N_I^\text{in}(i_k)\cap\tilde{N}_I^\text{in}(j_k)) \\
		\tilde{x}_{l}^{i_k}(k),& l\in \tilde{N}_I^\text{in}(i_k)\backslash (N_I^\text{in}(i_k)\cap\tilde{N}_I^\text{in}(j_k)).
		\end{cases}
		\end{eqnarray}
		For $i\in V,\,j\in\tilde{N}_I^\text{in}(i)$,
		\begin{equation}\label{other_inc_exc_I}
		\hat{x}_j^i(k)=\tilde{x}_j^i(k),\quad\forall i\neq i_k,\,\forall j\in\tilde{N}_I^\text{in}(i).
		\end{equation}
	\end{itemize}
	We suggest a compact form of gossip protocol by using a communication weight matrix $W^I(k)$. Let for player $i$,
	\begin{itemize}
		\item $m_i^\text{in}:=\text{deg}_{G_I}^\text{in}(i)+1$, where $\text{deg}_{G_I}^\text{in}(i)$ is the in-degree of vertex $i\in V$ in $G_I$,
		\item $m:=\sum_{i=1}^{N}m_i^\text{in}$,
		\item $B=[b_{ij}]_{i,j\in V}$ where	$b_{ij}=1$ if $j\in \tilde{N}_I^\text{in}(i)$ and $b_{ij}=0$, otherwise.
		\item $s_{ij}:=\sum_{l=1}^{j}B(i,l)+\delta_{i\neq 1}\sum_{r=1}^{i-1}m_r^\text{in}$, where $\delta_{i\neq 1} =1$ if $i\neq 1$ and $\delta_{i\neq 1} =0$, otherwise.
		\item $W^I(k):=I_m-\sum_{l\in(\tilde{N}_I^\text{in}(i_k)\cap\tilde{N}_I^\text{in}(j_k))}\frac{e_{s_{i_kl}}(e_{s_{i_kl}}-e_{s_{j_kl}})^T}{2}$,
		\vspace{-0.7cm}\begin{eqnarray}
		\label{WI(k)}
		\end{eqnarray}
		where $e_i\in\mathbb{R}^{m}$ is a unit vector. Note that $W^I(k)$ is different from the doubly stochastic one used in \cite{ssalehisadaghiani2016distributed}.
		\item $\tilde{x}(k):=\big[\tilde{x}^{1^T},\ldots,\tilde{x}^{N^T}\big]^T$: Stack vector of all temporary estimates of the players,
		\item $\bar{x}(k):=W^I(k)\tilde{x}(k)$: Intermediary variable.
	\end{itemize}
	Using the intermediary variable, one can construct the final estimates as follows:
	\begin{equation}\label{hat_be_bar}\hat{x}_{-i}^i(k)=[\bar{x}_{s_{ij}}(k)]_{j\in N_I^\text{in}(i)}.\end{equation}
	3- \textbf{\emph{Local Step:}} Player $i$ updates his action as follows:\\
	If $i=i_k$,
	\begin{eqnarray}
	\hspace{-0cm}x_i(k+1)\!=\!T_{\Omega_i}\!\Big[x_i(k)\!-\!\alpha_{k,i} \nabla_{x_i}J_i\big(x_i(k)\!,\![\bar{x}_{s_{ij}}(k)]_{j\in N_I^\text{in}(i)}\big)\!\Big],\nonumber
	\end{eqnarray}
	otherwise,
	\begin{equation}\label{local_step_I}
	x_i(k+1)=x_i(k),
	\end{equation} 
	where $T_{\Omega_i}:\mathbb{R}\rightarrow\Omega_i$ is an Euclidean projection and $\alpha_{k,i}$ is defined as in \eqref{diminish_step}.
	
	At this moment the updated actions are available for players to update their temporary estimates for every $i\in V,\,j\in\tilde{N}_I^\text{in}(i)$ as follows:
	\begin{equation}
	\label{temp_update_I}
	\tilde{x}_j^i(k+1)=\begin{cases} \bar{x}_{s_{ij}}(k),&\text{if }j\neq i \\ x_i(k+1),& \text{if }j=i.\end{cases}\end{equation}
	
	At this point, the players are ready to begin a new iteration from step 2.
	\begin{algorithm}
		\caption{}
		\begin{algorithmic}[1]
			\State \textbf{initialization} $\tilde{x}^i(0)\in\Omega^i\quad\forall i\in V$
			\For{$k=1,2,\ldots$ }
			\State \hspace{-0.5cm}$i_k\in V$ and $j_k\in N_C^\text{in}(i_k)$ communicate.
			\State \hspace{-0.5cm}$W^I\!(\!k\!)\!:=\!I_m\!-\!\sum_{l\in(\tilde{N}_I^\text{in}(i_k)\cap\tilde{N}_I^\text{in}(j_k))}\!\frac{e_{s_{i_kl}}\!(\!e_{s_{i_kl}}\!-\!e_{s_{j_kl}}\!)^T}{2}\!$.
			\State \hspace{-0.5cm}$\bar{x}(k)=W^I(k)\tilde{x}(k)$.
			\State \hspace{-0.5cm}$x_{i}(k\!+\!1)\!=\!T_{\Omega_{i}}[x_{i}(k)\!-\!\alpha_{k,{i}} \nabla_{x_i}J_{i}(x_{i}(k)\!,\![\bar{x}_{s_{ij}}(k)]_{j\in N_I^\text{in}(i)})]$
			if $i=i_k$,	$x_i(k+1)=x_i(k)$, otherwise.
			\State \hspace{-0.5cm}$\tilde{x}^i(k+1)=\bar{x}^i(k)+(x_i(k+1)-\bar{x}_i^i(k))e_i,\quad\forall i\in V$.
			\EndFor
		\end{algorithmic}
	\end{algorithm}
	\section{Convergence of the algorithm adapted to $G_I$} \label{convergence_diminish_interference}
	Similar to Section~\ref{convergence_diminish}, the convergence proof is split into two steps:
	\begin{enumerate}
		\item First, we prove almost sure convergence of $\tilde{x}(k)\subset\mathbb{R}^m$ to an average consensus which is shown to be the augmented average of all temporary estimate vectors. Let
		\begin{itemize}
			\item $m_i^\text{out}:=\text{deg}_{G_I}^\text{out}(i)+1$, where $\text{deg}_{G_I}^\text{out}(i)$ is the out-degree of vertex $i\in V$ in $G_I$,
			\item $1./\textbf{m}^\text{out}:=[\frac{1}{m_1^\text{out}},\ldots,\frac{1}{m_N^\text{out}}]^T$,
			\item $H:=[\sum_{i:1\in N_I^\text{in}(i)}e_{s_{i1}},\ldots,\sum_{i:N\in N_I^\text{in}(i)}e_{s_{iN}}]\in\mathbb{R}^{m\times N}$,
			\vspace{-0.75cm}
			\begin{eqnarray}
			\label{HI}
			\end{eqnarray}
			where $i:j\in N_I^\text{in}(i)$ is all $i$'s such that $j\in N_I^\text{in}(i)$.
		\end{itemize}
		The \emph{augmented average} of all temporary estimates is denoted by $Z^I(k)\in\mathbb{R}^m$ and defined as follows:
		\begin{equation}\label{ave_Z_I}
		Z^I(k):=H\text{diag}(1./\textbf{m}^\text{out})H^T\tilde{x}(k)\in\mathbb{R}^{m}.
		\end{equation} 
		\item Secondly, we prove almost sure convergence of the players’ actions toward the NE. %
	\end{enumerate}
	The convergence proof depends on some key properties of $W^I$ and $H$ given in Lemmas~\ref{lemma_ext_stoch}, \ref{gamma_less_1_I}.
	\begin{lemma}\label{lemma_ext_stoch}
		Let $W^I(k)$ and $H$ be defined in \eqref{WI(k)} and \eqref{HI}. Then, $W^I(k)H=H$.	Note that this can be interpreted as the generalized row stochastic property of $W^I(k)$.
	\end{lemma}
	\par{\emph{Proof}}. The proof is similar to the proof of Lemma~3 in \cite{salehisadaghiani2017adistribute} adapted for the different $W^I$ here. Using the definitions of $H$ and $W^I(k)$ \eqref{HI}, \eqref{WI(k)}, we expand $W^I(k)H$ as
	\begin{eqnarray}
	&&W^I(k)H=H-\frac{1}{2}\sum_{l\in(\tilde{N}_I^\text{in}(i_k)\cap\tilde{N}_I^\text{in}(j_k))}e_{s_{i_kl}}\nonumber\\
	&&.\bigg[\sum_{i:1\in N_I^\text{in}(i)}(e_{s_{i_kl}}-e_{s_{j_kl}})^Te_{s_{i1}},\ldots,\sum_{i:N\in N_I^\text{in}(i)}(e_{s_{i_kl}}-e_{s_{j_kl}})^Te_{s_{iN}}\bigg],
	\end{eqnarray}
	Note that $\sum_{i:j\in N_I^\text{in}(i)}(e_{s_{i_kl}}-e_{s_{j_kl}})^Te_{s_{ij}}=0$ for all $j\in V$ because $e_{s_{i_kl}}^Te_{s_{ij}}=1$ for $i=i_k$, $j=l$ and $e_{s_{i_kl}}^Te_{s_{ij}}=0$, otherwise. Similarly, $e_{s_{j_kl}}^Te_{s_{ij}}=1$ for $i=j_k$, $j=l$ and $e_{s_{i_kl}}^Te_{s_{ij}}=0$, otherwise. This completes the proof.
	$\hfill\blacksquare$
	
	Note that the generalized non-doubly stochasticity of $W^I(k)$ is translated into $H^TW^I(k)\neq H^T$.
	\begin{lemma}\label{gamma_less_1_I}
		Let $Q^I(k)\!:=\!W^I(k)-H\text{diag}(1./\textbf{m}^\text{out})H^TW^I(k)$ and\\ $\gamma^I=\lambda_{\max}\big(\mathbb{E}[Q^I(k)^TQ^I(k)]\big)$. Then $\gamma^I<1$.
	\end{lemma}
	\par{\emph{Proof}}. As suggested in \eqref{gamma_optimiz}, we employ the variational characterization of $\gamma$. Then,
	\begin{eqnarray}\label{gamma_optimiz_I}
	\hspace{-0.7cm}&&\gamma^I=\sup_{x\in\mathbb{R}^{m},\|x\|=1}x^T\mathbb{E}[Q^I(k)^TQ^I(k)]x\nonumber\\
	\hspace{-0.7cm}&&=\sup_{x\in\mathbb{R}^{m},\|x\|=1}x^T\mathbb{E}\Big[\Big(W^I(k)^T-W^I(k)^TH\text{diag}(1./\textbf{m}^\text{out}H^T)\Big)\nonumber\\
	\hspace{-0.7cm}&&.\Big(W^I(k)-H\text{diag}(1./\textbf{m}^\text{out})H^TW^I(k)\Big)\Big]\nonumber\\
	\hspace{-0.7cm}&&=\sup_{x\in\mathbb{R}^{m},\|x\|=1}x^T\mathbb{E}\Big[\Big(W^I(k)^TW^I(k)-W^I(k)^TH\text{diag}(1./\textbf{m}^\text{out})H^TW^I(k)\Big)\Big].
	\end{eqnarray} 
	For the last equality, we used $H^TH=\text{diag}(\textbf{m}^\text{out})$ which is straightforward to verify.
	We expand $\gamma^I$ and split the terms as follows (Let $l\in\tilde{N}_I^\text{in}(i_k)\cap\tilde{N}_I^\text{in}(j_k)$):
	\begin{eqnarray}\label{term1-term2_I}
	\hspace{-0.7cm}&&\gamma^I\!=\!\sup_{x}\! x^T\mathbb{E}\Big[\!\Big(\!\underbrace{I_m\!-\!\frac{1}{4}\sum_{l}(e_{s_{i_kl}}\!-\!e_{s_{j_kl}})e_{s_{i_kl}}^TH\text{diag}(\!1./\textbf{m}^\text{out}\!)}_{\text{Term}~1}\nonumber\\
	\hspace{-0.7cm}&&\underbrace{.H^T\sum_{l}e_{s_{i_kl}}(e_{s_{i_kl}}-e_{s_{j_kl}})^T}_{\text{Term}~1}\Big)-\Big(\underbrace{H\text{diag}(1./\textbf{m}^\text{out})H^T}_{\text{Term}~2}\nonumber\\
	\hspace{-0.7cm}&&\underbrace{-\frac{1}{4}\sum_{l}(e_{s_{i_kl}}-e_{s_{j_kl}})e_{s_{i_kl}}^T\sum_{l}e_{s_{i_kl}}(e_{s_{i_kl}}-e_{s_{j_kl}})^T}_{\text{Term}~2}\nonumber\\
	\hspace{-0.7cm}&&\underbrace{+\frac{1}{2}(I_m-H\text{diag}(1./\textbf{m}^\text{out})H^T)\sum_{l}e_{s_{i_kl}}(e_{s_{i_kl}}-e_{s_{j_kl}})^T}_{\text{Term}~2}\nonumber\\
	\hspace{-0.7cm}&&\underbrace{+\frac{1}{2}(\sum_{l}(e_{s_{i_kl}}-e_{s_{j_kl}})e_{s_{i_kl}}^T)(I_m-H\text{diag}(1./\textbf{m}^\text{out})H^T)}_{\text{Term}~2}\Big)\Big]x.\nonumber
	\end{eqnarray}
	We aim to prove that $x^T\mathbb{E}[\text{Term~1}]x\leq1$. Multiplying $x^T$ and $x$ into Term~1, we arrive at,
	\begin{eqnarray}
	\hspace{-0.7cm}&&x^T\mathbb{E}[\text{Term~1}]x=1-\frac{1}{4}\mathbb{E}\Big[\Big\|\text{diag}(1./\sqrt{\textbf{m}^\text{out}})H^T\sum_{l}e_{s_{i_kl}}(e_{s_{i_kl}}-e_{s_{j_kl}})^Tx\Big\|^2\Big]\leq1.
	\end{eqnarray}
	The equality holds for all $x$'s that satisfy $H^T\sum_{l}e_{s_{i_kl}}(e_{s_{i_kl}}-e_{s_{j_kl}})^Tx=0$.
	After a few manipulations, by the strong connectivity of $G_C$ (Remark~\ref{G_C_Strong}) for $i\in V$, $j\in N_C^\text{in}(i)$ and $l\in(\tilde{N}_I^\text{in}(i_k)\cap\tilde{N}_I^\text{in}(j_k))$ we obtain,
	\begin{equation}\label{equality_x_ha}
	x_{s_{il}}=x_{s_{jl}}.
	\end{equation} 
	To complete the proof we need to show $x^T\mathbb{E}[\text{Term~2}]x>0$ for all $x$'s satisfy \eqref{equality_x_ha} and $\|x\|=1$. After some manipulations we obtain,
	\begin{eqnarray}
	\hspace{-0.7cm}&&x^T\mathbb{E}[\text{Term~2}]x=x^TH\text{diag}(1./\textbf{m}^\text{out})H^Tx=\|\text{diag}(1./\sqrt{\textbf{m}^\text{out}})H^Tx\|^2\geq0
	\end{eqnarray} 
	The rest of the proof is straightforward by verifying that for all $x$'s which satisfy \eqref{equality_x_ha} and $\|x\|=1$, $H^Tx\neq0$.
	$\hfill\blacksquare$
	\begin{theorem}\label{consensus1_I}
		Let $\tilde{x}(k)$ be the stack vector with all temporary estimates of the players and $Z^I(k)$ be its average as in \eqref{ave_Z_I}. Let also $\alpha_{k,\text{max}}=\max_{i\in V}\alpha_{k,i}$. Then under Assumptions~\ref{G_I_Strongly_Connected}, \ref{connectivity_GTR}, \ref{assump}$^\prime$, the following hold.\\
		i) $\sum_{k=0}^{\infty}\alpha_{k,\text{max}}\|\tilde{x}(k)-Z^I(k)\|<\infty$,\qquad ii) $\sum_{k=0}^{\infty}\|\tilde{x}(k)-Z^I(k)\|^2<\infty$.
	\end{theorem}
	\par{\emph{Proof}}. The proof is based on Lemmas~\ref{lemma_ext_stoch},~\ref{gamma_less_1_I} and follows similar to the proof Theorem~1 in \cite{salehisadaghiani2017adistribute}.
	\begin{corollary}\label{remark_I}
		Let $z^I(k):=\text{diag}(1./\textbf{m}^\text{out})H^T\tilde{x}(k)\in\mathbb{R}^N$ be the average of all players' temporary estimates. Under Assumptions~\ref{G_I_Strongly_Connected}, \ref{connectivity_GTR}, \ref{assump}$^\prime$, the following hold for players' actions $x(k)$ and $\bar{x}(k)$:\\
		i) $\sum_{k=0}^{\infty}\alpha_{k,\text{max}}\|x(k)-z^I(k)\|<\infty$,\qquad ii) $\sum_{k=0}^{\infty}\|x(k)-z^I(k)\|^2<\infty$,\\
		iii) $\sum_{k=0}^{\infty}\mathbb{E}\Big[\|\bar{x}(k)-Z^I(k)\|^2\Big |\mathcal{M}_{k}\Big]<\infty$.
	\end{corollary}
	\par{\emph{Proof}}. The proof follows by taking into account $x(k)=[\tilde{x}_i^i(k)]_{i\in V}$, $Z^I(k)=Hz^I(k)$, $\bar{x}(k)=W^I(k)\tilde{x}(k)$ and also using Theorem~\ref{consensus1_I}.
	\begin{theorem}
		\label{Prop_algo_1_I}
		Let $x(k)$ and $x^*$ be all players' actions and the NE of $\mathcal{G}$, respectively. Under Assumptions~\ref{assump}$^\prime$-\ref{Lip_assump2}$^\prime$, \ref{G_I_Strongly_Connected}, \ref{connectivity_GTR}, the sequence $\{x(k)\}$ generated by the algorithm converges to $x^*$, almost surely.
		\vspace{-0.25cm}
	\end{theorem}
	\par{\emph{Proof}}. The proof uses Theorem~\ref{consensus1_I} and is similar to the proof of Theorem~2 in \cite{salehisadaghiani2017adistribute}. 
	\section{Simulation Results}
	\subsection{Social Media Behavior}
	In this example we aim to investigate social networking media for users' behavior. In such media like Facebook, Twitter and Instagram users are allowed to follow (or be friend with) the other users and post statuses, photos and videos or also share links and events. Depending on the type of social media, the way of communication is defined. For instance, in Instagram, friendship is defined unidirectional in a sense that either side could be only a follower and/or being followed.
	
	Recently, researchers at Microsoft have been studying the behavioral attitude of the users of Facebook as a giant and global network \cite{Forbes}. This study can be useful in many areas e.g. business (posting advertisements) and politics (posting for the purpose of presidential election campaign).
	
	Generating new status usually comes with the cost for the users such that if there is no benefit in posting status, the users don't bother to generate new ones. In any social media drawing others' attention is one of the most important motivation/stimulation to post status \cite{goel2012game}. Our objective is to find the optimal rate of posting status for each selfish user to draw more attention in his network. In the following, we make an information/attention model of a generic social media \cite{goel2012game} and define a way of communication between users ($G_C$) and an interference graph between them ($G_I$).
	
	Consider a social media network of $N$ users. Each user $i$ produces $x_i$ unit of information that the followers can see in their news feeds. The users' communication network is defined by a strongly connected digraph $G_C$ in which \raisebox{.5pt}{\textcircled{\raisebox{-.9pt} {$i$}}}$\rightarrow$\raisebox{0.5pt}{\textcircled{\raisebox{-.5pt} {$j$}}} means $j$ is a follower of $i$ or $j$ receives $x_i$ in his news feed. We also assume a strongly connected interference digraph $G_I$ to show the influence of the users on the others. We assume that each user $i$'s cost function is not only affected by the users he follows, but also by the users that his followers follow.
	
	The cost function of user $i$ is denoted by $J_i$ and consists of three parts:  
	\begin{enumerate}
		\item $C_i(x_i)$: A cost that user $i$ pays to produce $x_i$ unit of information.
		\begin{equation*}\label{facebook_Produce_Cost}
		C_i(x_i):=h_ix_i,\quad h_i>0.
		\end{equation*}
		\item $f_i^1(x)$: A differentiable, increasing and concave utility function of user $i$ from receiving information from his news feed with $f_i^1(\textbf{0})=0$.
		\begin{equation*}\label{facebook_information_utility}
		f_i^1(x):=L_i\sqrt{\sum_{j\in N_C^\text{in}(i)}q_{ji}x_j},\quad L_i>0,
		\end{equation*}
		where $q_{ji}$ represents follower $i$'s interest in user $j$'s information and $L_i$ is a user-specific parameter.\\
		\item $f_i^2(x)$: An incremental utility function that each user obtains from receiving attention in his network with $f_i^2(x)|_{x_i=0}=0$. Specifically, this function targets the amount of attention that each follower pays to the information of other users in his news feed.
		\begin{eqnarray}\label{facebook_attention_utility}
		\hspace{-1.4cm}&&f_i^2(x)=\sum_{l:i\in N_C^\text{in}(l)}\!L_l\Big(\sqrt{\!\sum_{j\in N_C^\text{in}(l)}q_{jl}x_j}-\sqrt{\!\sum_{j\in N_C^\text{in}(l)\backslash\{i\}}q_{jl}x_j}\Big).\nonumber
		\end{eqnarray}
	\end{enumerate}
	
	The total cost function for user $i$ is then $J_i(x)=C_i(x_i)-f_i^1(x)-f_i^2(x)$.
	For this example, we consider 5 users in the social media whose network of followers $G_C$ is given in Fig. 2. (a). From $G_C$ and taking $J_i$ into account, one can construct $G_I$ (Fig. 2. (b)) in a way that the interferences among users are specified.
	\begin{figure}
		\vspace{-0.87cm}
		\hspace{-6cm}
		\centering
		\includegraphics [scale=0.6]{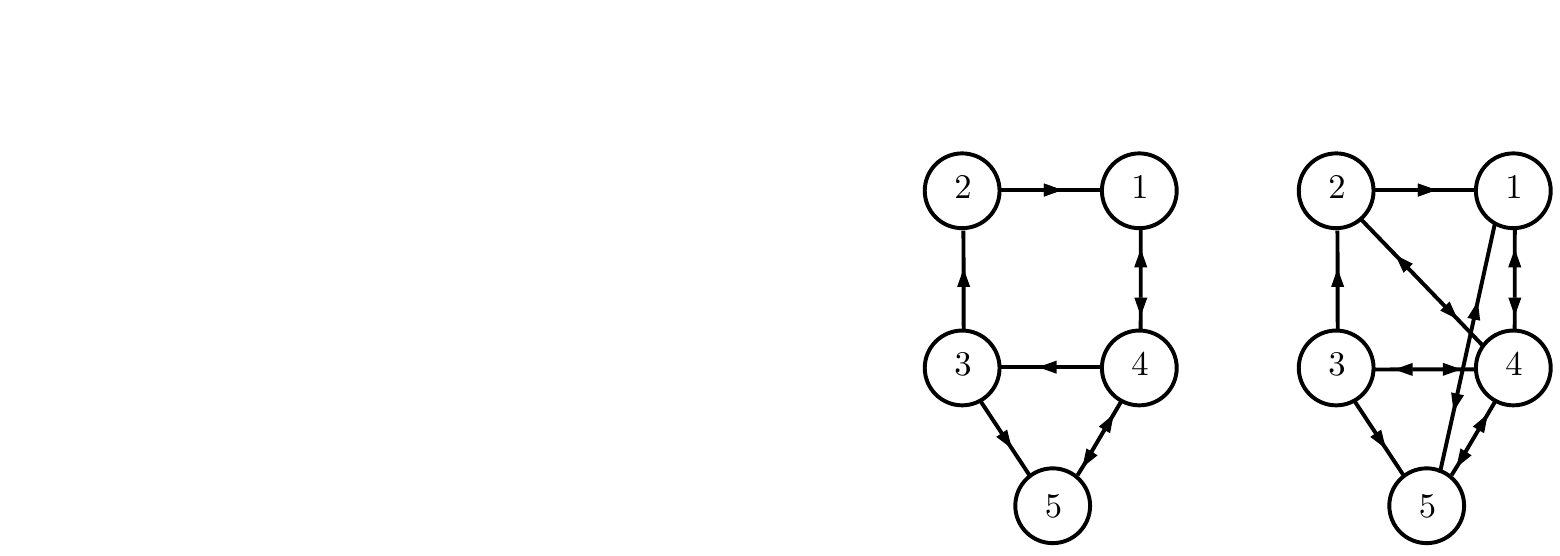}
		\label{fig11111}
		\caption{(a) $G_C$ (b) $G_I$.}
		\vspace{-0.6cm}
	\end{figure}  
	Note that this is a reverse process of the one discussed in Section~\ref{Section_Interference} because $G_C$ is given as the network of followers and $G_I$ is constructed from $G_C$. For the particular networks in Fig. 2, Assumptions~\ref{G_I_Strongly_Connected}, \ref{connectivity_GTR} hold. We then employ the algorithm in Section~\ref{asynch_interference} to find an NE of this game for $h_i=2$ and $L_i=1.5$ for $i\in V$, and $q_{41}=q_{45}=1.75$, $q_{32}=q_{43}=2$ and the rest of $q_{ij}=1$.
	\begin{figure}
		\vspace{-3.1cm}
		\hspace{-0.5cm}
		\centering
		\includegraphics [scale=0.4]{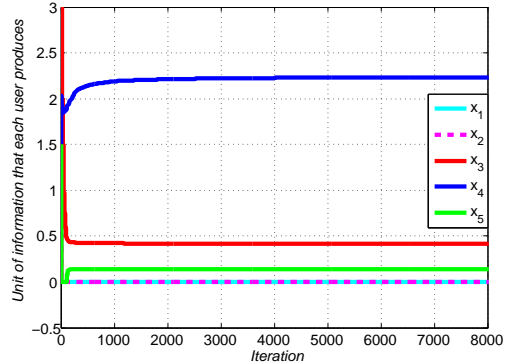}
		\label{fig11111}
		\vspace{-3cm}
		\caption{Convergence of the unit of information that each user produces to a NE over $G_C$.}
		\vspace{-0.6cm}
	\end{figure} 
	\subsection{Analysis}
	In this section we analyze the NE $x^*=[0,0,0.42,2.24,0.14]^T$. From $G_C$ in Fig. 2 (a), one can realize that user 4 has 3 followers (users 1, 3 and 5), user 3 has 2 followers (users 2 and 5) and the rest has only 1 follower. It is straightforward to predict that users 4 and 3 could draw more attentions due to their more number of followers which is end up having less cost. This let them to produce more information ($x_4^*\geq x_3^*\geq x_{j\in\{1,2,5\}}^*$). On the other hand, user 5 receives $x_4$ and $x_3$ from his news feed which ends up having greater payoff than users 1 and 2 from perceiving information. This is why $x_5^*\geq x_{j\in\{1,2\}}^*$.
	\section{Conclusions}
	We proposed an asynchronous gossip-based algorithm to find an NE of a networked game over a complete graph. Then, we extended our algorithm for the case of graphical games. We specified the locality of cost functions using an interference graph. Then, we provide a convergence proof to an NE of the game under an assumption on the communication graph.
	%
	%
	%
	%
	%
	%
	%
	%
	%
	%
	\bibliographystyle{splncs03}
	\bibliography{ref}
\end{document}